# Multi-Agent Modeling Using Intelligent Agents in the Game of Lerpa.

**Evan Hurwitz and Tshilidzi Marwala**
School of Electrical and Information Engineering
University of the Witwatersrand
Johannesburg, South Africa
laped@global.co.za

**Abstract –** *Game-theory has many limitations implicit in its application. By utilizing multi-agent modeling, it is possible to solve a number of problems that are unsolvable using traditional game-theory. In this paper reinforcement learning is applied to neural networks to create intelligent agents. Utilizing intelligent agents, intelligent virtual players learn from each other and their own rewards to play the game of Lerpa. These agents not only adapt to each other, but are even able to anticipate each other's reactions, and to "bluff" accordingly should the occasion arise. By pre-dealing specific hands, either to one player or to the whole table, one may "solve" the game, finding the best play to any given situation.*

## 1  Introduction

Current game analysis methods generally rely on the application of traditional *game-theory* to the system of interest [6]. While successful in simple systems/games, anything remotely complex requires the simplification of the said system to an approximation that can be handled by game-theory, with its not unsubstantial limitations [9]. An alternative approach is to rather analyze the game from within the *Multi-agent Modeling* (MAM) paradigm. While this approach traditionally utilizes simple agents [29], far too simplistic to handle any game of reasonable complexity, creating intelligent agents however, offers the possibility to "solve" these complex systems. The solution then becomes plausible without oversimplifying the system, as would be required in order to analyze them from a traditional game-theory perspective.

This paper explores the use of intelligent agents in a multi-agent system framework, in order to model and gain insight into a non-trivial, episodic card game with multiple players. The requirements for a multi-agent model will be specified, as will those for intelligent agents. Intelligent agents are then applied to the problem at hand, namely the card game, and results are observed and interpreted by utilizing sound engineering principles. All of this is presented with a view of evaluating the feasibility of applying intelligent agents within a multi-agent model framework, in order to solve complex games beyond the scope of traditional game-theory.

# 2  Game Theory

Game Theory is concerned with finding of the best, or *dominant* strategy (or strategies, in the case of multiple, equally successful strategies) in order to maximize a player's winnings within the constraints of a game's rule-set [19]. Game theory makes the following assumptions when finding these dominant strategies [11]:

- Each player has two or more well-specified choices or sequences of choices

- Every possible combination of choices leads to a well-specified end-state that terminates the game.

- A numerically meaningful payoff/reward is defined for each possible end-state.

- Each player has a *perfect knowledge* of the game and its opposition. *Perfect knowledge* implies that it knows the full rules of the game, as well as the payoffs of the other players.

- All decision-makers are rational. This implies that a player will always choose the action that yields the greatest payoff.

With these assumptions in mind, the game can then be analyzed from any given position by comparing the rewards of all possible strategy combinations, and then, by the last assumption, declaring that each player will choose the strategy with its own highest expected return [11].

## 2.1  Limitations of Game-Theory

As a result of the assumptions made when applying game theory, and of the methods themselves, the techniques developed have some implicit limitations [9]. A few of the more pressing limitations, which are not necessarily true of real-world systems, are as follows:

Game theory methods arrive at static solutions. These methods will deduce a solution, or *equilibrium point* for a given situation in a game. In many games, however, one will find that a solution changes or evolves as players learn the particular favored strategy of a player, and subsequently adapt to exploit the predictable behavior that results from playing only dominant strategies, and thus ultimately defeating the player utilizing the dominant strategy.

Real players are not always *rational*, as defined above. A player may display preferences, often seemingly at odds with statistically "best-play" strategies, which can change the odds within a game. A good strategy should account for this possibility and exploit it, rather than ignore it. This problem is referred to in economic and game-theory circles as the "trembling hand".

Game theory cannot handle more than two to three players. Due to dimensionality issues, game theory cannot be used to analyze games with a large number of players without simplifying the game to two- or three-player games, game complexity having the final word on the player limit.

Game theory can only be applied to relatively simple games. Once again as a result of dimensionality issues, complex games have too many states to be effectively analyzed using traditional game theory methods.

In order to be analyzed, many complex games are simplified by dividing the players into grouped "camps", effectively reducing multi-player games into two- or three-player games. Likewise the rules are often simplified, in order to similarly reduce the dimensionality of a game. While these simplifications may allow analysis to proceed, they also change the fundamental nature of many games, rendering the analysis flawed by virtue of examining an essentially dissimilar game.

These limitations within game-theory prompt the investigation of an alternative method of analyzing more complex games. This paper investigates the option of utilizing *Multi-Agent Modeling*, using intelligent agents, to analyze a complex game.

## 3   Multi-Agent Modelling

In its simplest form, multi-agent modeling involves breaking a system up into its component features, and modeling those components in order to model the overall system [10]. Central to this methodology is the notion of *emergent behavior*, that is, that the simple interactions between agents produce complex results [8]. These results are often far more complex than the agents that gave rise to them [8].

### 3.1   Emergent Behavior

Emergent behavior is so pivotal to the understanding and utilization of multi-agent modeling, that a brief elaboration becomes necessary. While not, strictly speaking, an instance of MAM, John Conway's *game of artificial life* provides an excellent illustration of emergent behavior, and the ramifications thereof [11]. In Conway's game, an MxN grid of squares (often infinite) each contains one binary value, being either *alive* or *dead*. In each iteration of the game, a *dead* square will become alive if exactly three adjacent squares are also alive, and an *alive* square will *die* if there are less than two adjacent *living* squares, or if there are more than three adjacent *living* squares, as depicted in Figure 1.

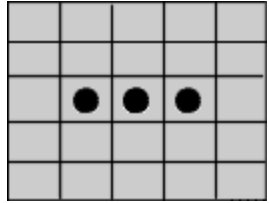 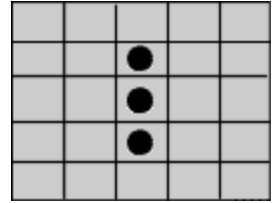

(a)                                                                                                      (b)

Figure 1. A single iteration of *Life*

In the next iteration, the two outermost living squares in Figure 1(a) will die since each has only one living neighbor, and the two squares above and below the centre living square will come to life, as each has exactly three living neighbors, resulting in the situation depicted in Figure 1(b). As one can see, the rules are incredibly simple, but the consequences of these rules, i.e. the *Emergent Behavior*, are far from simple. Figure 2 shows a simple-looking shape, commonly referred to as a *glider* [11].

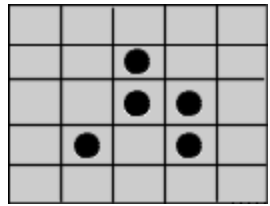

Figure 2. Simple Glider

This shape continues to stably propagate itself at a forty-five degree angle within the game. In contrast, the even simpler-looking shape in Figure 3, known as an *r-pentamino*, produces an explosion of shapes and patterns that continually change and mutate, only becoming predictable after 1103 iterations [11].

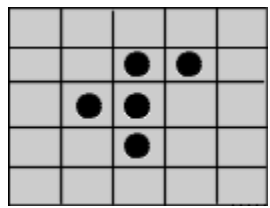

Figure 3. R-Pentamino

It is this same complex behavior, emerging from interacting components following simple rules, that lies at the heart of MAM's promise, and making it such a powerful tool [8].

## 3.2 Advantages of Multi-Agent Modeling

An agent-based model holds many advantages over standard analytical techniques, which are traditionally mathematical or statistical in nature [5]. Specifically, some of the advantages offered by an agent-based model are as follows [14]:

- Agents are far simpler to model than the overall system they comprise, and hence the system becomes easier to model.

- The emergent behavior resulting from the agent interactions implies that systems too complex to be traditionally modeled can now be tackled, since the complexity of the system need not be explicitly modeled.

- Large systems with heterogeneous agents can be easily handled within a multi-agent system, while this is incredibly difficult to cater for using traditional mathematics, which would make the often unrealistic demand that the components be homogenous.

## 3.3 Weaknesses/Limitations

While MAM has definite advantages, it is not without weaknesses. Since the emergent behavior is arrived at empirically, and is not deterministic, it is difficult to state with any degree of certainty as to why a certain outcome has been arrived at [26]. Similarly, since emergent behavior is often unexpected [8], it can be difficult to ascertain whether the multi-agent system (MAS) is incorrectly modeling the system in question. Thus, validation of the model becomes an important aspect of any MAS.

## 3.4 MAM Applications

Multi-agent modeling lends itself to a number of applications. The following are some of the more common applications of multi-agent modeling:

### 3.4.1 Swarm Theory

Multi-agent modeling is utilized for the development of *Swarm Theory* based systems [4]. These systems utilize many simple agents, and attempt to design simple individual rules that will allow the agents to work together to achieve a larger, common goal, much in the same manner that a swarm of ants will work together to collect food for the colony. These systems depend on the engineer's ability to predict the (often unexpected) emergent behavior of the system for given agent behavior.

### 3.4.2 Complexity Modelling

Multi-agent modeling is well-suited to the task of *complexity modeling* [25]. Complexity modeling refers to modeling complex systems that are often too complex to be explicitly modeled [25]. The usage of representative agents allows for the emergent behavior of the

MAM to model the complexity within the system, rather than the said complexity being explicitly modeled by the engineer [25]. Essentially, the complexity is contained by the *interactions* between the agents, and between the agents and the system, rather than the traditional, and often insufficient, mathematical models previously used [25].

### 3.4.3 Economics

Fundamentally an application of complexity modeling, multi-agent modeling can be applied to economic systems [7]. This discipline, known as Applied Computational Economics (ACE), applies a bottom-up approach to modeling an economic system, rather than the traditional top-down approach, which requires full system specification and then component decomposition [7]. In order to verify ACE system veracity, the ACE model is required to reproduce known results empirically. Once this has been accomplished, the same system can then be used to predict the results of unknown situations, allowing for better forecasting and policy decision-making.

### 3.4.4 Social Sciences

Many attempts have been made to model social phenomena, with varying degrees of success [15]. Since social systems, by definition, involve the interaction of autonomous entities [27], multi-agent modeling offers an ideal methodology for modeling such systems [15]. The foundations of such applications have already been laid, with the groundwork being solutions to such problems as the *standing ovation* problem [15].

## 3.5 Making a Multi-agent Model

In order to create a multi-agent model, the smaller components that comprise the system must be specified [24]. These smaller components need to be fully modeled, so as to become the *agents* that are at the heart of the modeling technique [24]. Each agent must be capable of making decisions (often dictated by a rule-set), and these decisions may involve incomplete knowledge of its environment [24]. The agents also need to be able to receive information from their environments, and depending on the type of system being modeled, sometimes communicate with other agents [24]. The environment itself needs to be able to adjudicate the interactions between agents, but at no stage needs to be able to determine the overall ramifications of these interactions. Instead, the overall result will become apparent empirically, taking advantage of the emergent behavior of the multi-agent system to handle the complexity modeling [24].

# 4 Intelligent Agents

Artificial intelligence (A.I.) can only truly be considered worthy of the name when the system in question is capable of learning on its own, without having an expert *teacher* available to point out correct behavior. This leads directly into the paradigm of *reinforcement learning* [23]. Most *reinforcement learning* techniques explored utilize lookup-table system representations, or linear function approximators, which severely hinder the scope of learning available to the artificial intelligence system. One notable exception is the work on *TD-Gammon*, in which he successfully applied the TD($\lambda$) reinforcement learning algorithm to train a neural network, with staggeringly successful results [11]. Following attempts to emulate this work have, however, been met with failure due to the extreme difficulties of combining back-propagation with TD($\lambda$). Some methods for overcoming these problems have been explored, allowing for the combination of these versatile techniques in order to create an intelligent agent.

## 4.1 What is Intelligence?

In order to create intelligent agents, it is necessary to first define intelligence, so that we may critically evaluate whether the agent created meets these criteria. In order to be considered intelligent, an entity must be capable of the following [20]:

- The entity must be able to learn.

- The entity must be able to learn from its own inferences, without being taught.

- The entity must be capable of drawing conclusions from incomplete data, based on its own knowledge.

- The entity must be able to re-evaluate its own knowledge, and adapt if necessary.

Should an agent meet these requirements, it can then be considered intelligent. In order to meet these requirements, the agents will learn within the *reinforcement learning* paradigm.

## 4.2 Reinforcement Learning

Reinforcement learning involves the training of an artificial intelligence system by trial-and-error, reflecting the same manner of learning that living beings exhibit [23]. Reinforcement learning is very well suited to episodic tasks [21], and as such is highly appropriate in the field of game-playing, where many episodes are encountered before a final result is reached, and an A.I. system is required to evaluate the value of each possible move long before a final result is achieved. This methodology allows for online learning, and also eliminates the need for expert knowledge [23].

### 4.2.1 Rewards and Returns

Any artificial intelligence system requires some sort of goal or target to strive towards [28]. In the case of reinforcement learning, there are two such quantities that need to be defined, namely *rewards* and *returns* [23]. A reward is defined to be the numerical value attributed to an individual state, while a return is the final cumulative rewards that are returned at the end of the sequence [23]. The return need not necessarily be simply summed, although this is the most common method [23]. An example of this process can be seen below in Figure 4, where an arbitrary Markov process is illustrated with rewards given at each step, and a final return at the end. This specific example is that of a *Random walk* problem.

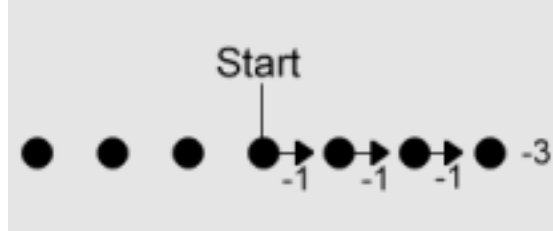

Figure 4. Random walk rewards and returns.

An A.I. system utilizing reinforcement learning must learn to predict its expected return at each stage, hence enabling it to make a decision that has a lower initial reward than other options, but maximizing its future return. This can be likened to making a sacrifice in chess, where the initial loss of material is accepted for the future gains it brings.

### 4.2.2  Exploitation vs Exploration

An A.I. system learning by reinforcement learning learns only through its own experiences [23]. In order to maximize its rewards, and hence its final return, the system needs to experiment with decisions not yet tried, even though it may perceive them to be inferior to tried-and-tested decisions [23]. This attempting of new approaches is termed *exploration*, while the utilization of gained knowledge to maximize returns is termed *exploitation* [23]. A constant dilemma that must be traded off in reinforcement learning is that of the choice between exploration and exploitation. One simple approach is the *ε-greedy* approach, where the system is *greedy*, i.e. attempts to exploit, with probability ε. Hence the system will explore with probability 1- ε [23].

### 4.2.3  TD (λ)

One common method of training a reinforcement learning system is to use the TD (λ) (Temporal Difference) algorithm to update one's value estimates [22][23]. This algorithm is specifically designed for use with episodic tasks, being an adaptation of the common Widdrow-Hoff learning rule [21]. In this algorithm, the parameters or *weights w* to be altered are updated by equation (1) [21].

$$\Delta w = \alpha (P_{t+1} - P_t) \sum_{k=1}^{t} \lambda^{t-k} \nabla w P_k \qquad (1)$$

The prediction $P_{t+1}$ is used as a target for finding the error in prediction $P_t$, allowing the update rule to be computed incrementally, rather than waiting for the end of the sequence before any learning can take place [22]. Parameters α and λ are the learning rate and weight-decay parameters, respectively.

The prediction P can be made in a number of different methods, ranging from a simple lookup table to complex function approximators [21]. Owing to the nature of the task at hand, a function approximator with a high degree of flexibility is required, since the agents must be capable of drawing any logical links they see fit, and not be limited by our choice of function

## 5   Neural Networks

It is necessary to understand the workings and advantages of neural networks to appreciate the task of applying them in the reinforcement learning paradigm. It is likewise important to fully grasp the implications of reinforcement learning, and the break they represent from the more traditional supervised learning paradigm.

### 5.1   Neural network architecture

The fundamental building-blocks of neural networks are *neurons* [28]. These neurons are simply a multiple-input, single-output mathematical function [28]. Each neuron has a number of *weights* connecting it to inputs from a previous layer, which are then added together, possibly with a *bias*, the result of which is then passed into the neuron's *activation function* [28]. The activation function is a function that represents the way in which the neural network "thinks". Different activation functions lend themselves to different problem types, ranging from yes-or-no decisions to linear and nonlinear mathematical relationships. Each *layer* of a neural network is comprised of a finite number of neurons. A network may consist of any number of layers, and each layer may contain any number of neurons [28]. When a neural network is run, each neuron in each consecutive layer sums its inputs and multiplies each input by its respective weight, and then treats the weighted sum as an input to its activation function. The output will then be passed on as an input to the next layer, and so on until the final output layer is reached. Hence the input data is passed through a network of neurons in order to arrive at an output. Figure 5 illustrates an interconnected network, with 2 input neurons, three hidden layer neurons, and two output neurons. The hidden layer and output layer neurons can all have any of the possible activation functions. This type of neural network is referred to as a *multi-layer perceptron* [28], and while not the only configuration of neural network, it is the most widely used configuration for regression-type problems [18].

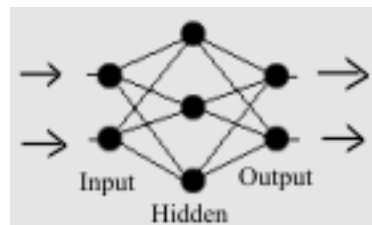

Figure 5. Sample connectionist network

## 5.2 Neural network properties

Neural networks have various properties that can be utilized and exploited to aid in the solving of numerous problems. Some of the properties that are relevant to this particular problem are detailed below.

### 5.2.1 Universal approximators

Multi-layer feed-forward neural networks have been proven to be universal approximators [28]. By this one refers to the fact that a feed-forward neural network with nonlinear activation functions of appropriate size can approximate any function of interest to an arbitrary degree of accuracy [28]. This is contingent upon sufficient training data and training being supplied.

### 5.2.2 Neural Networks Can Generalize

By approximating a nonlinear function from its inputs, the neural network can learn to approximate a function [28]. In doing so, it can also infer the correct output from inputs that it has never seen, by inferring the answer from similar inputs that it has seen. This property is known as *generalization* [28]. As long as the inputs received are within the ranges of the training inputs, this property will hold [28].

### 5.2.3 Neural Networks Recognize Patterns

Neural networks are often required to match large input/output sets to each other, and these sets are often 'noisy' or even incomplete [28]. In order to achieve this matching, the network learns to recognize patterns in the data sets rather than fixate on the answers themselves [28]. This enables a network to 'see' through the data points and respond to the underlying pattern instead. This is an extended benefit of the generalization property.

## 5.3 Training

Training of neural networks is accomplished through the use of an appropriate algorithm [16]. The two main types of training algorithms employed are back-propagation and batch updating algorithms [16]. Many algorithms exist, all with their own unique advantages and disadvantages. Commonly used back-propagation algorithms are Steepest Descent and Scaled Conjugate Gradient training methods [16], while a commonly used batch updating algorithm is the Quasi-Newton training algorithm [16]. All of these algorithms are gradient-based algorithms for multivariable optimization, which are preferable to evolutionary methods due to their guaranteed convergence, even though global optimality cannot be guaranteed [16]. Regardless of the variation, the fundamental idea behind back-propagation methods is that weights are updated, via the optimization method selected, from the last neuron layer back to the first [16], whereas batch updating algorithms update all weights simultaneously, making for more complex computations, although fewer required iterations [16].

## 5.4 Generalization

Neural networks are capable of generalizing to situations that they have not in fact been trained on, providing that they have been trained over an encompassing range of inputs [3]. In other words, a network trained on input values of 1; 4; 7; and 10 will be able to present an accurate answer to an input of 8 since the network has been trained with values both greater and smaller than 8. It will however, struggle to present a correct answer to an input of 11 since the highest input training value was only 10. The ability to generalize is at the heart of the over-fitting/overtraining issue [3]. It is generally accepted that overtraining is a myth, provided that the number of hidden neurons is correct and that the training data is complete. However, since satisfying both of these conditions is nontrivial, the problem of overtraining remains a stark reality. Overtraining leads to the more specific problem of over-fitting, wherein a network fits too closely to data that may be incorrect, often fitting the function to the noise rather than the underlying pattern [3]. A visual example of over-fitting and in contrast good generalization is presented in Figure 6.

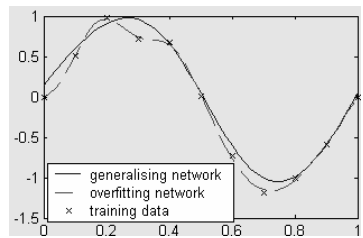

Figure 6. Over-fitted Vs generalising Networks

## 5.5 Suitability of Neural Networks for an Intelligent MAM Agent

A neural network would seem to be an exceptionally good, and obvious choice for use as a function approximator for the agent's predictor. The fact that it is a universal approximator means that the agent is capable of making any inference it deems fit to link the inputs to its observed values, thus not constraining the agent to limited human/expert knowledge, which a linear function approximator would. A comprehensive database lookup-table could provide a similar result, but it has a number of difficulties that prohibit such a measure. The first problem would be that of *dimensionality*, i.e. that there are simply too many possible states to encode, often more in total than even modern computer memory can handle. On the contrary, the compact nature of a neural network allows for all of these states to be captured using minimal memory. In addition, the training time for any sort of convergence would be extremely slow with a lookup-table method, again due to its size. Differing methods for generalizing using eligibility traces [18] do exist, but all would impose an artificial linking between states that may not necessarily exist. A neural network's generalization property, however, allows for training of similar states without imposing any such constraints on the budding Artificial Intelligence's freedom.

With these advantages in mind, the next problem is observed to be the act of training the network, since the standard algorithms are of no use in the reinforcement learning paradigm. The standard gradient-based algorithms require input/output data sets upon which to train [16], which for the following reasons our agent will not have access to:

- Datasets would require a known *correct* answer, which would prevent the agent from finding better answers than those currently accepted.

- Datasets can grow outdated swiftly, especially in a competitive gaming situation, where one's opponents can form an integral part of the game's optimum strategy.

For these reasons, the TD($\lambda$) algorithm has been selected for use, based largely on the success of the Tessauro TD-Gammon program.

### 5.6 Jumping the hurdles

In order to implement the TD($\lambda$) algorithm for training neural networks, it is better to first tackle a known, smaller problem. By first tackling such a problem, one can identify and solve the prevailing issues within the application of the algorithm, without them being clouded by issues pertaining to the more complex system. For these reasons, the problem of tic-tac-toe has been tackled, with an agent created to play against itself, learning to play the game as it plays. The simplicity of the game makes analysis of the agents easy, and the finite, solved nature of the game means that one can easily determine whether the agent has made a good or a bad decision. This last facet will not be true of larger games, where the agent could easily become better than the standard by which it is judged, and is thus very important to establish now, when evaluating the methodology.

## 6 Tic Tac Toe

The game of *Tic Tac Toe*, or *noughts and crosses*, is played on a 3x3 grid, with players taking alternate turns to fill an empty spot [12][13]. The first player places a 'O', and the second player places a 'X' whenever it is that respective player's turn [13]. If a player manages to get three of his mark in a row, he wins the game [13]. If all 9 squares are filled without a winner, the game is a draw [13]. The game was simulated in MATLAB, with a simple matrix representation of the board.

### 6.1 Player evaluation

The A.I. system, or *player*, needs to be evaluated in order to compare different players, who have each learned using a different method of learning. In order to evaluate a player's performance, ten different positions are set up, each with well-defined correct moves. Using this test-bed, each player can be scored out of ten, giving a measure of the level of play each player has achieved. Also important is the speed of convergence – i.e. how fast does each respective player reach its own maximum level of play.

## 6.2 TD(λ) for backpropogation

In order to train a neural network, equation (1) needs to be adapted for use with the back-propagation algorithm [23]. The adaptation, without derivation, is as follows [23]:

$$w_{ij}^{t+1} = w_{ij}^t + \alpha \sum_{K \in O} \left( P_K^{t+1} - P_K^t \right) e_{ijk}^t \quad (2)$$

where the eligabilities are:

$$e_{ijk}^{t+1} = \lambda e_{ijk}^t + \delta_{kj}^{t+1} y_i^{t+1} \quad (3)$$

and δ is calculated by recursive back-propagation

$$\delta_{ki}^t = \frac{\partial P_k^t}{\partial S_i^t} \quad (4)$$

as such, the TD(λ) algorithm can be implemented to update the weights of a neural network [23].

## 6.3 Stability issues

The TD(λ) algorithm has proven stability for linear functions [22]. A multi-layer neural network, however, is non-linear [16], and the TD(λ) algorithm can become unstable in some instances [23] [22]. The instability can arise in both the actual weights and in the predictions [23] [22]. In order to prevent instability, a number of steps can be taken, the end result of which is in most cases to limit the degree of variation in the outputs, so as to keep the error signal small to avoid instability.

### 6.3.1 Input/Output Representation

The inputs to, and outputs from, a typical A.I. system are usually represented as real or integer values. This is not optimal for TD(λ) learning, as the values have too much variation. Far safer is to keep the representations in binary form, accepting the dimensionality trade-off (a function of the number of inputs into the network, and hence larger using a binary representation) as a fair price to ensure a far higher degree of stability. Specifically in the case of the outputs, this ensures that the output error of the system can never be more than 1 for any single output, thus keeping the mean error to within marginally stable bounds. For the problem of the Tic Tac Toe game, the input to the network is an 18-bit binary string, with the first 9 bits representing a possible placed 'o' in each square, and the second 9 bits representing a '1' in each square. The output of the network is a 3-bit string, representing an 'o' win, a draw and an 'x' win respectively.

### 6.3.2 Activation Functions

As shown in Section 2, there are many possible activation functions that can be used for the neural network. While it is tempting to utilize activation functions that have a large scope in order to maximize the versatility of the network, it proves far safer to use an activation function that is limited to an upper bound of 1, and a lower bound of zero. A commonly used activation function of this sort is the sigmoidal activation function, having the form of [16]:

$$f(x) = \frac{1}{1+e^{-x}} \tag{5}$$

This function is nonlinear, allowing for the freedom of approximation required of a neural network, and limiting the upper and lower bounds as recommended above. While this activation function is commonly used as a middle-layer activation function, it is unusual as an output layer activation function. In this manner, instability is further discouraged.

### 6.3.3 Learning Rate

As the size of the error has a direct effect on the stability of the learning system, parameters that directly effect the error signal also have an effect on the said stability. For this reason, the size of the learning rate α needs to be kept low, with experimental results showing that values between 0.1 and 0.3 prove safe, while higher values tend to become unstable, and lower values simply impart too little real learning to be of any value.

### 6.3.4 Hybrid Stability Measures

In order to compare relative stability, the percentage chance of becoming unstable has been empirically noted, based upon experimental results. Regardless of each individual technique presented, it is the combination of these techniques that allows for better stability guarantees. While no individual method presented gives greater than a 60% stability guarantee (that is, 60% chance to be stable given a 100-game training run), the combination of all of the above measures results in a much better 98% probability of being stable, with minor tweaking of the learning rate parameter solving the event of instability occurring at unusual instances.

## 6.4 The Players

All of the players are trained using an ε-greedy policy, with the value of ε = 0.1. i.e. for each possible position the player has a 10% chance of selecting a random move, while having a 90% chance of selecting whichever move it deems to be the best move. This selection is done by determining all of the legal moves available, and then finding the positions that would result from each possible move. These positions are sequentially presented to the player, who then rates each resultant position, in order to find the best resultant position. It obviously follows that whichever move leads to the most favored position is the apparently best move, and the choice of the player for a greedy policy. The training of each player is accomplished via *self-play*, wherein the player evaluates and chooses moves for both sides, learning from its own experiences as it discovers errors on its own. This learning is continued until no discernable improvement occurs.

As a benchmark, randomly initialized networks were able to correctly solve between 1 and 2 of the posed problems, beyond which one can say genuine learning has indeed taken place, and is not simply random chance.

## 6.5 Player #1 – Simple TD($\lambda$)

Player #1 learned to play the game using a simple TD($\lambda$) back-propagation learning algorithm. This proved to be very fast, allowing for many thousands of games to be played out in a very short period of time. The level of play achieved using this method was however not particularly inspiring, achieving play capable of solving no more than 5 of the 10 problems posed in the rating system. The problem that is encountered by player #1 is that the learning done after each final input, the input with the game result, gets undone by the learning of the intermediate steps of the next game. While in concept the learning should be swifter due to utilizing the knowledge gained, the system ends up working at cross-purposes against itself, since it struggles to build its initial knowledge base, due to the generalization of the neural network which is not present in more traditional reinforcement learning arrangements.

## 6.6 Player #2 – Historical Database Learning

In this instance, the player learns by recording each position and its corresponding target, and storing the pair in a database. Duplicate input data sets and their corresponding targets are removed, based on the principle that more recent data is more accurate, since more learning has been done when making the more recent predictions. This database is then used to train the network in the traditional supervised learning manner. A problem encountered early on with this method is that early predictions have zero knowledge base, and are therefore usually incorrect. The retaining of this information in the database therefore taints the training data, and is thus undesirable. A solution to this problem is to limit the size of the database, replacing old data with new data once the size limit is reached, thus keeping the database recent. This methodology trains slower than that employed to train player #1, making long training runs less feasible than for TD($\lambda$) learning. The play level of this method is the lowest of those examined, able to solve only four of the ten proposed problems. Nonetheless, the approach does show promise for generating an initial knowledge base from which to work with more advanced methods.

## 6.7 Player #3 – Fact/Opinion DB Learning

Building on the promise of Player #2, a more sophisticated database approach can be taken. If one takes into account the manner of the training set generation, one notes that most of the targets in the database are no more than *opinions* – targets generated by estimates of the next step, as seen in equation (1) – while relatively few data points are in fact *facts* – targets generated by viewing the end result of the game. In order to avoid this problem, the database can be split into two sub-databases, with one holding facts, and the other holding opinions. Varying the sizes, and the relative sizes, of these two sub-databases can then allow the engineer to decide how much credence the system should give to fact versus opinion. This method proved far more successful than Player #2, successfully completing 6 of the 10 problems posed by the rating system. It's speed of convergence is comparable to that of Player #2.

## 6.8 Player #4 – Widdrow-Hoff based DB Learning

In this instance, a very similar approach to that of Player #2 is taken, with one important distinction: Instead of estimating a target at each move, the game is played out to completion with a static player. After each game finishes, the player then adds all of the positions encountered into its database, with the final result being the target of each position. This means that no *opinions* can ever enter into the training, which trades off speed of convergence for supposedly higher accuracy. This method is not optimal, as it loses one of the primary advantages of reinforcement learning, namely that of being able to incorporate current learning into its own learning, hence speeding up the learning process. Unsurprisingly, this method trains with the same speed as the other database methods, but takes far longer to converge. It achieves a similar level of play as does Player #1, being able to solve 5 of the posed problems.

## 6.9 Player #5 - Hybrid Fact/Opinion DB TD($\lambda$) Learning

The logical extension to the previous players is to hybridise the most successful players in order to compensate for the failings of each. Player #5 thus utilizes the Fact/Opinion database learning in order to build an initial knowledge base from which to learn, and then proceeds to learn from thence using the TD($\lambda$) approach of Player #1. This proves more successful, since the intrinsic flaw in player #1's methodology lies in its inability to efficiently create a knowledge base, and the database method of player #3 creates that knowledge base from which to learn. Player #5 begins its learning with the expected sluggishness of database methods, but then learns much faster once it begins to learn using the TD($\lambda$) approach. Player #5 managed to successfully solve seven of the ten problems once trained to convergence. The problem of unlearning learned information is still apparent in Player #5, but is largely mitigated by the generation of the initial knowledge base.

## 6.10 Player Comparisons

As is illustrated in Figure 7, the hybrid method learns to play at the strongest level of all of the methods presented. Due to the drastic differences in speed and computational power requirements, it is preferable to stay away from database-based methods, and it is thus worth noting that only the fact/opinion database method arrives at a stronger level of play than the simple TD($\lambda$)-trained player #1, and that this methodology can easily be incorporated into a TD($\lambda$) learning system, which produces the far more promising player #5.

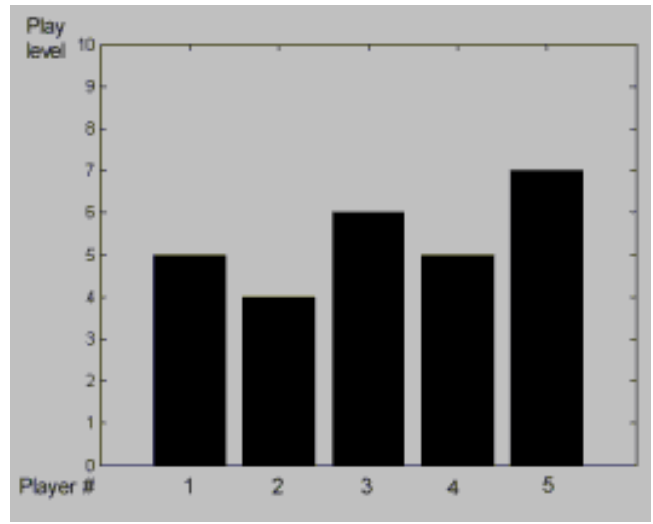

Figure 7. Relative player strengths

The fact that after a short knowledge-base generation sequence the hybrid system uses the highly efficient TD($\lambda$) approach makes it a faster and more reliable learning system than the other methods presented. As can be seen in Figure 7, however, there is still a greater level of play strength that should be achievable in this simple game, and that has been limited by the unlearning error seen in Players #1 and #5.

While no perfect result is achieved, the primary goal of learning and adapting is successful. The agents have been shown to learn on their own, without tutoring. They can infer from past knowledge to make estimates of unknown situations, and can adapt to changing situations. Thus the agents have satisfied all four requirements necessary to be considered "intelligent agents".

# 7   Lerpa

The card game being modeled is the game of Lerpa. While not a well-known game, its rules suit the purposes of this research exceptionally well, making it an ideal testbed application for intelligent agent. The rules of the game first need to be elaborated upon, in order to grasp the implications of the results obtained. Thus, the rules for Lerpa now follow.

The game of *Lerpa* is played with a standard deck of cards, with the exception that all of the 8s, 9s and 10s are removed from the deck. The cards are valued from greatest- to least-valued from ace down to 2, with the exception that the 7 is valued higher than a king, but lower than an ace, making it the second most valuable card in a suit. At the end of dealing the hand, during which each player is dealt three cards, the dealer has the choice of *dealing himself in* – which entails flipping his last card over, unseen up until this point, which then declares which suit is the *trump suit*. Should he elect not to do this, he then flips the next card in the deck to determine the trump suit. Regardless, once trumps are determined, the players then take it in turns, going clockwise from the dealer's left, to elect whether or not to play the hand (to *knock*), or to drop out of the hand, referred to as *folding* (If the Dealer has *dealt himself in*, as described above, he is then automatically required to play the hand). Once all players have chosen, the players that have elected to play then play the hand, with the player to the dealer's left playing the first card. Once this card has been played, players must then play *in suit* – in other words, if a heart is played, they must play a heart if they have one. If they have none of the required suit, they may play a trump, which will win the trick unless another player plays a higher trump. The highest card played will win the trick (with all trumps valued higher than any other card) and the winner of the trick will lead the first card in the next trick. At any point in a hand, if a player has the Ace of trumps and can legally play it, he is then required to do so. The true risk in the game comes from the betting, which occurs as follows:

At the beginning of the round, the dealer pays the table 3 of whatever the basic betting denomination is (referred to usually as 'chips'). At the end of the hand, the chips are divided up proportionately between the winners, i.e. if you win two tricks, you will receive two thirds of whatever is in the pot. However, if you stayed in, but did not win any tricks, you are said to have been *Lerpa'd*, and are then required to match whatever was in the pot for the next hand, effectively costing you the pot. It is in the evaluation of this risk that most of the true skill in *Lerpa* lies.

## 8 Lerpa MAM

As with any optimization system, very careful consideration needs to be taken with regards to how the system is structured, since the implications of these decisions can often result in unintentional assumptions made by the system created. With this in mind, the Lerpa MAS has been designed to allow the maximum amount of freedom to the system, while also allowing for generalization and swift convergence in order to allow the intelligent agents to interact unimpeded by human assumptions, intended or otherwise.

### 8.1 System Overview

The game is, for this model, going to be played by four players. Each of these players will interact with each other indirectly, by interacting directly with the *table*, which is their shared environment, as depicted in Figure 8.

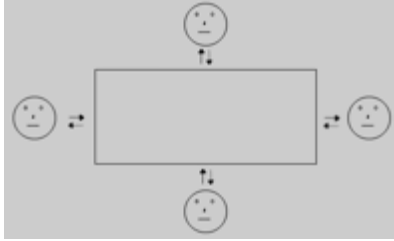

Figure 8. System interactions.

Over the course of a single hand, an agent will be required to make three decisions, once at each interactive stage of the game. These three decision-making stages are:

- Whether to play the hand, or drop (*knock* or *fold*)

- Which card to play first

- Which card to play second

Since there is no decision to be made at the final card, the hand can be said to be effectively finished from the agent's perspective after it has played its second card (or indeed after the first decision should the agent fold). Following on the TD($\lambda$) algorithm, each agent will update its own neural network at each stage, using its own predictions as a reward function, only receiving a true reward after its final decision has been made. This decision making process is illustrated below, in Figure 9.

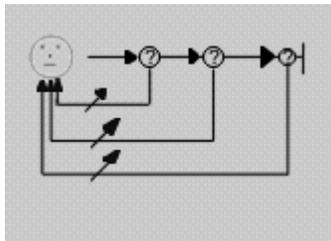

Figure 9. Agent learning scheme

With each agent implemented as described, they can now interact with each other through their shared environment, and will continuously learn upon each interaction and its consequent result.

Each hand played will be viewed as an independent, stochastic event, and as such only information about the current hand will be available to the agent, who will have to draw on its own learned knowledge base to draw deductions from rather than from previous hands.

## 8.2 Agent AI Design

A number of decisions need to be made in order to implement the agent AI effectively and efficiently. The type of learning to be implemented needs to be chosen, as well as the neural network architecture. Special attention needs to be paid to the design of the inputs to the neural network, as these determine what the agent can 'see' at any given point. This will also determine what assumptions, if any, are implicitly made by the agent, and hence cannot be taken lightly. Lastly, this will determine the dimensionality of the network, which directly affects the learning rate of the network, and hence must obviously be minimized.

### 8.2.1 Input Parameter Design

In order to design the input stage of the agent's neural network, one must first determine all that the network may need to know at any given decision-making stage. All inputs, in order to optimize stability, are structured as binary-encoded inputs. When making its first decision, the agent needs to know its own cards, which agents have stayed in or folded, and which agents are still to decide. It is necessary for the agent to be able to match specific agents to their specific actions, as this will allow for an agent to learn a particular opponent's characteristics, something impossible to do if it can only see a number of players in or out. Similarly, the agent's own cards must be specified fully, allowing the agent to draw its own conclusions about each card's relative value. It is also necessary to tell the agent which suit has been designated the trumps suit, but a more elegant method has been found to handle that information, as will be seen shortly. Figure 10 below illustrates the initial information required by the network.

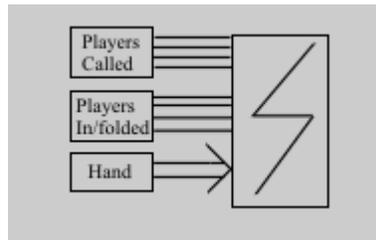

Figure 10. Basic input structure.

The agent's hand needs to be explicitly described, and the obvious solution is to encode the cards exactly, i.e. four suits, and ten numbers in each suit, giving forty possibilities for each card. A quick glimpse at the number of options available shows that a raw encoding style provides a sizeable problem of dimensionality, since an encoded hand can be one of $40^3$ possible hands (in actuality, only $^{40}P_3$ hands could be selected, since cards cannot be repeated, but the raw encoding scheme would in fact allow for repeated cards, and hence $40^3$ options would be available). The first thing to notice is that only a single deck of cards is being used, hence no card can ever be repeated in a hand. Acting on this principle, consistent ordering of the hand means that the base dimensionality of the hand is greatly reduced, since it is now combinations of cards that are represented, instead of permutations. The number of combinations now represented is $^{40}C_3$. This seemingly small change from $^nP_r$ to $^nC_r$ reduces the dimensionality of the representation by a factor of r!, which in this case is a factor of 6. Furthermore, the representation of cards as belonging to discrete suits is not optimal either, since the game places no particular value on any suit by its own virtue, but rather by virtue of which suit is the trump suit. For this reason, an alternate encoding scheme has been determined, rating the 'suits' based upon the makeup of the agent's hand, rather than four arbitrary suits. The suits are encoded as belonging to one of the following groups, or new "suits":

- Trump suit

- Suit agent has multiple cards in (not trumps)

- Suit in agent's highest singleton

- Suit in agent's second-highest singleton

- Suit in agent's third-highest singleton

This allows for a much more efficient description of the agent's hand, greatly improving the dimensionality of the inputs, and hence the learning rate of the agents. These five options are encoded in a binary format, for stability purposes, and hence three binary inputs are required to represent the suits. To represent the card's number, ten discrete values must be represented, hence requiring four binary inputs to represent the card's value. Thus a card in an agent's hand is represented by seven binary inputs, as depicted in Figure 11.

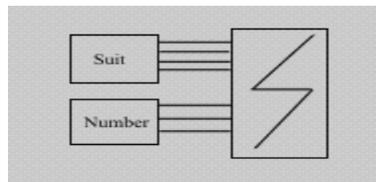

Figure 11. Agent card input structure

Next must be considered the information required in order to make decisions two and three. For both of these decisions, the cards that have already been played, if any, are necessary to know in order to make an intelligent decision as to the correct next card to play. For the second decision, it is also plausible that knowledge of who has won a trick would be important. The most cards that can ever be played before a decision must be made is seven, and since the table after a card is played is used to evaluate and update the network, it is necessary to represent eight played cards. Once again, however, simply utilizing the obvious encoding method is not necessarily the most efficient method. The actual values of the cards played are not necessarily important, only their values relative to the cards in the agent's hand. As such, the values can be represented as one of the following, with respect to the cards of the same suit in the agent's hand:

- Higher than the card/cards in the agent's hand

- Higher than the agent's second-highest card

- Higher than the agent's third-highest card

- Lower than any of the agent's cards

- Member of a void suit (value is immaterial)

Another suit is now relevant for representation of the played cards, namely a void suit – a suit in which the agent has no cards. Lastly, a number is necessary to handle the special case of the Ace of trumps, since its unique rules mean that strategies are possible to develop based on whether it has or has not been played. The now six suits available still only require three binary inputs to represent, and the six number groupings now reduce the value representations from four binary inputs to three binary inputs, once again reducing the dimensionality of the input system.

With all of these inputs specified, the agent now has available all of the information required to draw its own conclusions and create its own strategies, without human-imposed assumptions affecting its "thought" patterns.

### 8.2.2  Network Architecture Design

With the inputs now specified, the hidden and output layers need to be designed. For the output neurons, these need to represent the prediction P that the network is making. A single hand has one of five possible outcomes, all of which need to be catered for. These possible outcomes are:

- The agent wins all three tricks, winning 3 chips.

- The agent wins two tricks, winning 2 chips.

- The agent wins one trick, winning 1 chip.

- The agent wins zero tricks, losing 3 chips.

- The agent elects to fold, winning no tricks, but losing no chips.

This can be seen as a set of options, namely [3 2 1 0 -3]. While it may seem tempting to output the result as one continuous output, there are two compelling reasons for breaking these up into binary outputs. The first of these is in order to optimize stability, as elaborated upon in Section 5. The second reason is that these are discrete events, and a continuous representation would cover the range of [-3 0], which does not in fact exist. The binary inputs then specified are:

- $P(O = 3)$

- $P(O = 2)$

- $P(O = 1)$

- $P(O = -3)$

With a low probability of all four catering to folding, winning and losing no chips. Consequently, the agent's predicted return is:

$$P = 3A + 2B + C - 3D \tag{6}$$

where

$$A = P(O = 3) \tag{7}$$

$$B = P(O = 2) \tag{8}$$

$$C = P(O = 1) \tag{9}$$

$$D = P(O = -3) \tag{10}$$

The internal structure of the neural network is consistent with that upon which the stability optimization in Section 5 was done, using equation (5) for the hidden layer activation function. Since a high degree of freedom is required, a high number of hidden neurons is required, and thus fifty have been used. This number is iteratively achieved, trading off training speed versus performance. The output neurons are linear functions, since they represent not binary effects, but rather a continuous probability of particular binary outcomes.

### 8.2.3 Agent Decision-Making

With its own predictor specified, the agent is now equipped to make decisions when playing. These decisions are made by predicting the return of the resultant situation arising from each legal choice it can make. An ε-greedy policy is then used to determine whether the agent will choose the most promising option, or whether it will explore the result of the less appealing option. In this way, the agent will be able to trade off exploration versus exploitation.

## 9 The Intelligent Model

With each agent implemented as described above, and interacting with each other as specified in section eight, we can now perform the desired task, namely that of utilizing a multi-agent model to analyze the given game, and develop strategies that may "solve" the game given differing circumstances.

### 9.1 Agent Learning Verification

In order for the model to have any validity, one must establish that the agents do indeed learn as they were designed to do. In order to verify the learning of the agents, a single intelligent agent was created, and placed at a table with three 'stupid' agents. These 'stupid' agents always stay in the game, and choose a random choice whenever called upon to make a decision. The results show quite conclusively that the intelligent agent soon learns to consistently outperform its opponents, as shown in Figure 12.

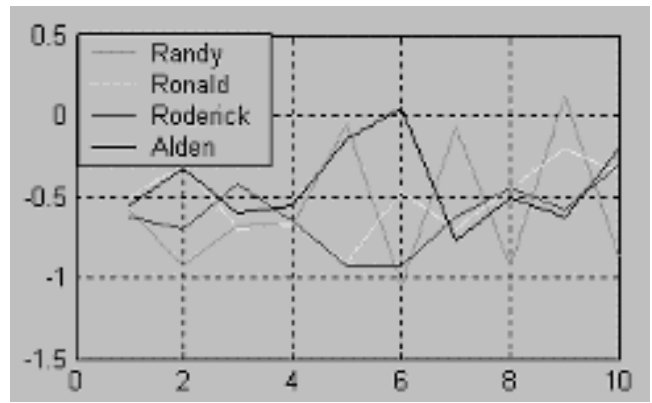

Figure 12 gent performance, averaged over 40 hands

The agents named Randy, Roderick and Ronald use random decision-making, while AIden has the TD($\lambda$) AI system implemented. The results have been averaged over 40 hands, in order to be more viewable, and to also allow for the random nature of cards being dealt. As can be seen, AIden is consistently performing better than its counterparts, and continues to learn the game as it plays.

### 9.1.1 Cowardice

In the learning phase of the abovementioned intelligent agent, an interesting and somewhat enlightening problem arises. When initially learning, the agent does not in fact continue to learn. Instead, the agent quickly determines that it is losing chips, and decides that it is better off not playing, and keeping its chips! This is illustrated in Figure 13.

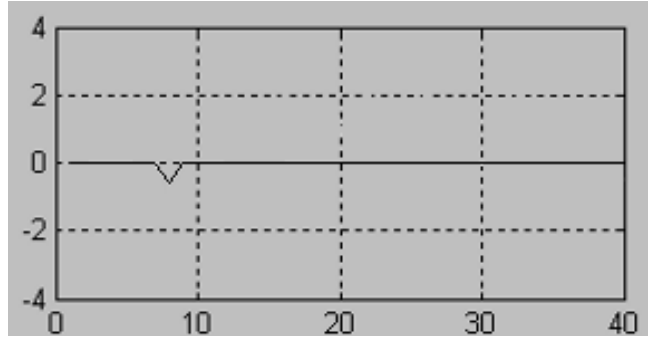

Figure 13. Agent cowardice. Averaged over 5 hands

As can be seen, AIden quickly decides that the risks are too great, and does not play in any hands initially. After forty hands, AIden decides to play a few hands, and when they go badly, gets scared off for good. This is a result of the penalizing nature of the game, since bad play can easily mean one loses a full three chips, and since the surplus of lost chips is nor carried over in this simulation, a bad player loses chips regularly. While insightful, a cowardly agent is not of any particular use, and hence the agent must be given enough 'courage' to play, and hence learn the game. In order to do this, one option is to increase the value of ε for the ε-greedy policy, but this makes the agent far too much like a random player without any intelligence. A more successful, and sensible solution is to force the agent to play when it knows nothing, until such a stage as it seems prepared to play. This was done by forcing AIden to play the first 200 hands it had ever seen, and thereafter leave AIden to his own devices.

## 9.2  Parameter Optimisation

A number of parameters need to be optimized, in order to optimize the learning of the agents. These parameters are the learning-rate α, the memory parameter λ and the exploration parameter ε. The multi-agent system provides a perfect environment for this testing, since four different parameter combinations can be tested competitively. By setting different agents to different combinations, and allowing them to play against each other for an extended period of time (number of hands), one can iteratively find the parameter combinations that achieve the best results, and are hence the optimum learning parameters. Figure 14 shows the results of one such test, illustrating a definite 'winner', whose parameters were then used for the rest of the multi-agent modeling. It is also worth noting that as soon as the dominant agent begins to lose, it adapts its play to remain competitive with its less effective opponents. This is evidenced at points 10 and 30 on the graph (games number 300 and 900, since the graph is averaged over 30 hands) where one can see the dominant agent begin to lose, and then begins to perform well once again.

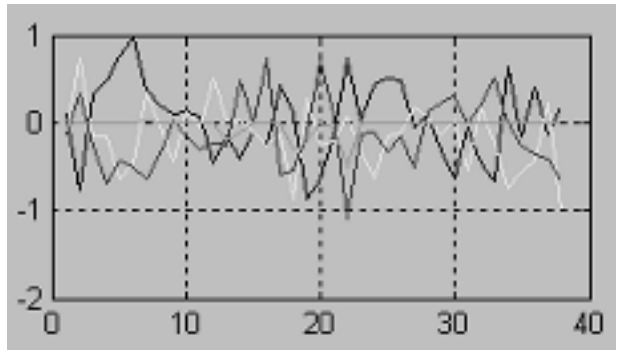

Figure 14. Competitive agent parameter optimisation. Averaged over 30 hands.

Surprisingly enough, the parameters that yielded the most competitive results were α = 0.1; λ = 0.1 and ε = 0.01. while the ε value is not particularly surprising, the relatively low α and λ values are not exactly intuitive. What they amount to is a degree of temperance, since higher values would mean learning a large amount from any given hand, effectively over-reacting when they may have played well, and simply have fallen afoul of bad luck.

## 9.3 MAS Learning Patterns

With all of the agents learning in the same manner, it is noteworthy that the overall rewards they obtain are far better than those obtained by the random agents, and even by the intelligent agent that was playing against the random agents. A sample of these results is depicted in Figure 15.

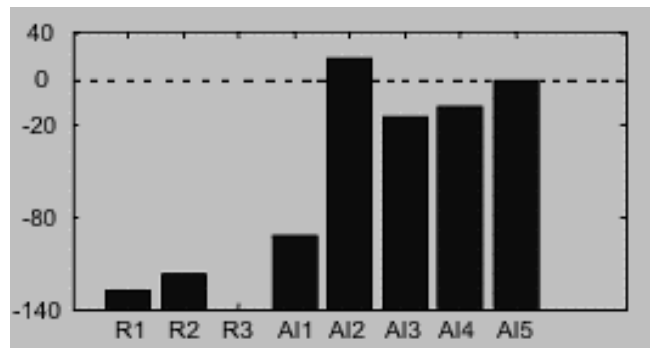

Figure 15. Comparative returns over 200 hands.

R1 to R3 are the Random agents, while AI1 is the intelligent agent playing against the random agents. AI2 to AI5 depict intelligent agents playing against each other. As can be seen, the agents learn far better when playing against intelligent opponents, an attribute that is in fact mirrored in human competitive learning [1]. The agents with better experience tend to fold bad hands, and hence lose far fewer chips than the intelligent agent playing against unpredictable opponents.

## 9.4 Agent Adaptation

In order to ascertain whether the agents in fact adapt to each other or not, the agents were given pre-dealt hands, and required to play them against each other repeatedly. The results of one such experiment, illustrated in Figure 16, shows how an agent learns from its own mistake, and once certain of it, changes its play, adapting in order to gain a better return from the hand. The mistakes it sees are its low returns, returns of -3 to be precise. At one point, the winning player obviously decides to explore, giving some false hope to the losing agent, but then quickly continues to exploit his advantage. Eventually, at game #25, the losing agent gives up, adapting his play to suit the losing situation in which he finds himself. Figure 16 also illustrates the progression of the agents and the adaptation described.

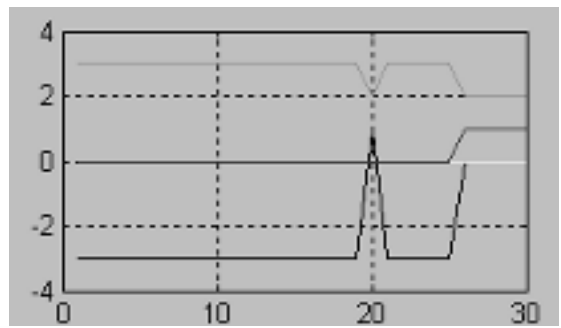

Figure 16. Adaptive agent behaviour

## 9.5 Strategy analysis

The agents have been shown to successfully learn to play the game, and to adapt to each other's play in order to maximize their own rewards. These agents form the pillars of the multi-agent model, which can now be used to analyze, and attempt to 'solve' the game. Since the game has a nontrivial degree of complexity, situations within the game are to be solved, considering each situation a sub-game of the overall game. The first, and most obvious type of analysis is a static analysis, in which all of the hands are pre-dealt. This system can be said to have stabilized when the results and the playout become constant, with all agents content to play the hand out in the same manner, each deciding that nothing better can be achieved. This is akin to Game Theory's "static equilibrium".

## 9.6 Bluffing

A bluff is an action, usually in the context of a card game, that misrepresents one's cards with the intent of causing one's opponents to drop theirs (i.e. to fold their hand). There are two opposing schools of thought regarding bluffing [2]. One school claims that bluffing is purely psychological, while the other maintains that a bluff is a purely statistical act, and therefore no less sensible than any other strategy [2]. Astoundingly enough, the intelligent agents do in fact learn to bluff! A classic example is illustrated in Figure 17, which depicts a hand in which bluffing was evidenced.

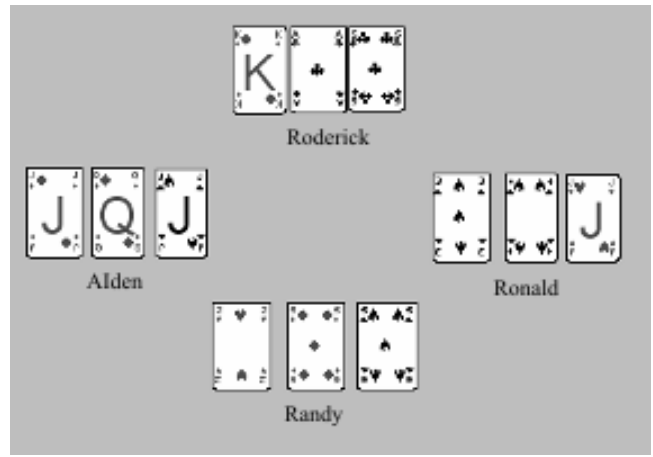

Figure 17. Agent bluffing

In the above hand, Randy is the first caller, and diamonds have been declared trumps. Randy's hand is not particularly impressive, having only one low trump, and two low supporting cards. Still, he has the lead, and a trump could become a trick, although his risks are high for minimal reward. Nonetheless, Randy chooses to play this hand. Ronald, having nothing to speak of, unsurprisingly folds. Roderick, on the other hand, has a very good hand. One high trump, and an outside ace. However, with one still to call, and Randy already representing a strong hand by playing, Roderick chooses to fold. AIden, whose hand is very strong with two high trumps and an outside jack, plays the hand. When the hand is played repeatedly, Randy eventually chooses not to play, since he loses all three to AIden. Instantly, Roderick chooses to play the hand, indicating that the bluff was successful, that it chased a player out of the hand! Depending on which of the schools of thought regarding bluffing one follows this astonishing result leads us to one of two possible conclusions. If, like the authors, one maintains that bluffing is simply playing the odds, making the odds for one's opponent unfavorable by representing a strong hand, then this result shows that the agents learn each other's patterns well enough to factor their opponent's strategies into the game evaluation, something Game Theory does a very poor job of. Should one follow the theory that bluffing is purely psychological, then the only conclusion that can be reached from this result is that the agents have in fact developed their own 'psyches', their own personalities which can then be exploited. Regardless of which option tickles the reader, the fact remains that agents have been shown to learn, on their own and without external prompting, to bluff! [12][17]

## 9.7 Deeper Strategy Analysis

While the strategy analysis already presented is useful, it is not truly practical, since for many applications one may have access to no more information than a single agent has. For such situations, one can perform a different simulation, pre-dealing only a single agent's hand. In this way, the agent will be testing its strategy against a dynamic opposition, being forced to value its hand on its own merit and not based on static results. Once the agent performs constantly with its hand, the resulting strategy will be the dominant strategy. To be said to be constantly performing, the agent should average zero or greater than zero returns when playing out according to its arrived at strategy, since negative returns would indicate playing with a weak hand that should be folded. In this way, one can determine the correct play with any given hand, effectively 'solving' the game.

## 9.8 Personality Profiling

While Game Theory stagnates in the assumption that all players are rational, there is no such limit strangling a multi-agent model using intelligent agents. While this may not seem important on the surface, it does in fact extend the usefulness of the model far beyond that of standard Game Theory. Many poker players of reasonable skill complain bitterly about playing against beginners, bemoaning the fact that the beginners do not play as they should, as a "rational" player would, and thus throw the better players off their game. A good player, however, should not be limited to assuming rationality from his opponents, but rather should identify his opponents' characteristics and exploit their weaknesses. In order to perform this task, one needs to create "personality" types in agents, this while sounding somewhat daunting, is in fact a rather simple task. All that is required is to modify the reward function for an agent (equation #6) to reflect the personality type to be modeled. Two personality types were created, namely an *aggressive* personality and a *conservative* personality. The aggressive agent uses the reward function:

$$P = 3A + 2B + C - 2D \tag{11}$$

While the conservative agent uses the reward function:

$$P = 3A + 2B + C - 4D \tag{12}$$

Where the symbols have the following meanings: P denotes the expected outcome of the hand; A denotes the probability of winning three tricks; B denotes the probability of winning two tricks; C denotes the probability of winning one trick; and D denotes the probability of winning zero tricks.

As can be seen, in this case both agents modify the coefficient of the D term in order to skew their world view. This is certainly not the only manner in which the reward function can be modified in order to reflect a personality, but is the most obvious, since the D term represents the "risk" that the agent sees within a hand. Using the above modifications, the previously detailed strategy analysis techniques can be performed on agents with distinctive personalities. The static analysis of comes to the same results as with rational players, due to the unchanging nature of the problem, while the dynamic analysis yields more interesting results. Dynamic strategy analysis finds different dominant strategies when playing against these profiled personalities. More specifically, the aggressive player is not considered as dangerous when playing in a hand, while the conservative player is treated with the utmost of respect, since he only plays the best of hands.

## 10 Conclusions

Mutli-agent modeling using intelligent agents allows one to analyze and solve games that traditional game theory struggles to handle. By utilizing reinforcement learning, and the TD($\lambda$) algorithm in particular, adaptive, intelligent agents can be created to interact within the multi-agent system. While these agents will learn against both intelligent and non-intelligent agents, they learn far faster against better, more intelligent agents, achieving a higher standard of play in a shorter period of time. They also continually adapt to each other's play styles, finding dynamic equilibria within the game. The system can be used to determine "best play" strategy for any given hand, in any specific scenario. Moreover, bluffing in the game was shown to be in all likelihood a natural strategic development, an act of "playing the odds", rather than the traditional view of being a psychological play. Through modifying the reward signals received by agents, it was also shown that personality types can be modeled, allowing for hands to be "solved" taking into account non-rational opponents, a feature sorely lacking in traditional game-theory. Finally, input sensitivity analysis was found to be unsuccessful in analyzing the decision-making process of the agents, by reason of the encoding scheme.